\documentclass[11pt,english]{article}
\usepackage{mathptmx}

\usepackage[T1]{fontenc}
\usepackage[latin9]{inputenc}
\usepackage{geometry}
\usepackage{float}
\usepackage{amsthm}
\usepackage{amsmath}
\usepackage{graphicx}
\usepackage{esint}

\makeatletter
\theoremstyle{plain}
\newtheorem{thm}{\protect\theoremname}
  \theoremstyle{definition}
  \newtheorem{defn}[thm]{\protect\definitionname}

\makeatother

\usepackage{babel}
  \providecommand{\definitionname}{Definition}
\providecommand{\theoremname}{Theorem}

\begin{document}

\title{The Synchrosqueezing transform for instantaneous spectral analysis}

\author{Gaurav Thakur}

\date{April 11, 2014}
\maketitle
\begin{abstract}
The Synchrosqueezing transform is a time-frequency analysis method
that can decompose complex signals into time-varying oscillatory components.
It is a form of time-frequency reassignment that is both sparse and
invertible, allowing for the recovery of the signal. This article
presents an overview of the theory and stability properties of Synchrosqueezing,
as well as applications of the technique to topics in cardiology,
climate science and economics. 
\end{abstract}

\section{Introduction}

The Synchrosqueezing transform is a time-frequency analysis method
that can characterize signals with time-varying oscillatory properties.
It is designed to analyze and decompose signals of the form

\begin{equation}
f(t)=\sum_{k=1}^{K}A_{k}(t)e^{2\pi i\phi_{k}(t)},\label{AMFM}
\end{equation}

where the $A_{k}$ and $\phi_{k}$ are time-varying amplitude and
phase functions respectively. The goal is to recover the \textit{instantaneous
frequencies} (IFs) $\{\phi_{k}'\}_{1\leq k\leq K}$ and the oscillatory
components $\{A_{k}e^{2\pi i\phi_{k}}\}_{1\leq k\leq K}$. Signals
of the form (\ref{AMFM}) arise in numerous scientific and engineering
applications%
\footnote{Such signals are often called ``nonstationary'' in these domains,
although this terminology is not related to its meaning for random
processes.%
} but are not well represented in a traditional Fourier basis, where
the individual elements of the basis fail to capture localized oscillations
in the components $\{A_{k}e^{2\pi i\phi_{k}}\}$. Standard time-frequency
methods such as the short-time Fourier transform (STFT) and the continuous
wavelet transform (CWT) are often used to analyze such signals, but
do not take advantage of any sparsity of the form (\ref{AMFM}) in
the signal and incur a tradeoff in time-frequency resolution \cite{Da92,Fl99}.
Synchrosqueezing is a variant of time-frequency reassignment (TFR),
a class of techniques that apply a nonlinear post-processing mapping
to a conventional STFT or CWT plot. The mapping is designed to ``push''
the energy in an STFT closer to its most prominent frequencies, resulting
in a sparse and concentrated time-frequency representation of the
signal \cite{AFL13,Fl02}. However, traditional TFR methods result
in a loss of information from the underlying transform and cannot
be used to recover the original signal, and also often involve heuristics
that are difficult to justify rigorously.\\

Synchrosqueezing combines the localization and sparsity properties
of TFR with the invertibility of a traditional time-frequency transform,
and is robust to a variety of disturbances in the signal. The main
concepts behind Synchrosqueezing were originally introduced in the
mid-1990s for audio signal analysis \cite{DM96}, but it has received
much closer attention in recent years, with an extensive mathematical
theory developed in \cite{DLW11} and \cite{TW11}. Unlike traditional
TFR, Synchrosqueezing performs the post-processing mapping only in
the frequency direction and does so in a manner that preserves the
total energy of the signal $f$, allowing for the decomposition of
the signal into the components $\{A_{k}e^{2\pi i\phi_{k}}\}$. This
article provides a concise survey of the Synchrosqueezing methodology
and its associated theory, and also discusses real-world applications
in several different domains where the technique has provided new
insights.

\section{The Synchrosqueezing process}

The Synchrosqueezing transform was originally developed in \cite{DLW11}
and \cite{DM96} in terms of the CWT. We choose a (complex) mother
wavelet $\psi$ such that the Fourier transform $\hat{\psi}$ has
strictly positive support and satisfies the standard admissibility
condition $\int_{0}^{\infty}z^{-1}\hat{\psi}(z)dz<\infty$ \cite{Da92}.
The CWT $W_{\psi}f(a,t)$ at the scale $a$ and time shift $t$ is
then given by
\begin{equation}
W_{\psi}f(a,t)=a^{-1/2}\int_{-\infty}^{\infty}f(u)\overline{\psi\left(\frac{u-t}{a}\right)}du.\label{CWT}
\end{equation}

We then take the \textit{phase transform} $\omega f(a,b)$, defined
as the derivative of the complex phase of $W_{\psi}f$,
\begin{equation}
\omega f(a,t)=\frac{\frac{\partial}{\partial t}W_{\psi}f(a,t)}{2\pi iW_{\psi}f(a,t)}.\label{PhaseTransform}
\end{equation}

Intuitively, this nonlinear operator can be thought of as removing
the influence of $\psi$ from the CWT and ``encoding'' the localized
frequency information we want. The key step is to consider the \textit{CWT
Synchrosqueezing transform},
\begin{equation}
S_{\epsilon}^{\delta,M}f(t,\eta)=\int_{\{(a,t):a\in[M^{-1},M],|W_{\psi}f(a,t)|>\epsilon\}}a^{-3/2}W_{\psi}f(a,t)\frac{1}{\delta}h\left(\frac{\eta-\omega f(a,t)}{\delta}\right)da\label{SST}
\end{equation}
for a test function $h\in C_{0}^{\infty}$, a sufficiently large parameter
$M$, and sufficiently small $\delta>0$ and $\epsilon>0$. The motivation
for (\ref{SST}) is that it is a smoothed out approximation to 
\[
Sf(t,\eta)=\int_{\{(a,t):\eta=\omega f(a,t)\}}a^{-3/2}W_{\psi}f(a,t)da,
\]
or in other words, a partial inversion of the CWT that is only taken
over the level curves of the phase transform $\omega f$ and ignores
the rest of the time-scale plane $(a,t)$. This localization process
allows us to recover the components $A_{k}e^{2\pi i\phi_{k}}$ more
accurately than inverting the CWT over the entire time-scale plane.
Alternatively, the mapping $W_{\psi}f(a,t)\to Sf(t,\eta)$ can be
thought of as a reassignment operation that squeezes energy from the
scales $a$ into IFs $\eta$ centered on the level curves of $\omega f$,
but leaves the total energy in $W_{\psi}f(a,t)$ at each time $t$
unchanged. For appropriate signals $f$, the energy in the Synchrosqueezing
transform $S_{\epsilon}^{\delta,M}f(b,\eta)$ is concentrated precisely
around the IF curves $\{\phi_{k}'(t)\}$. Finally, once $S_{\epsilon}^{\delta,M}f$
is computed, we can recover each of the components by completing the
inversion of the CWT and integrating over small bands around each
IF curve,
\begin{equation}
R_{k,\epsilon}^{\delta,M}f(t)=\frac{1}{\int_{0}^{\infty}\frac{\hat{\psi}(z)}{z}dz}\int_{|\eta-\phi_{k}'(t)|<\epsilon}S_{\epsilon}^{\delta,M}f(t,\eta)d\eta.\label{Inversion}
\end{equation}

Under certain conditions, it can be shown that $R_{k,\epsilon}^{\delta,M}f(t)\approx A_{k}(t)e^{2\pi i\phi_{k}(t)}$.
In practice, an additional, intermediate step is needed to identify
the integration bands in (\ref{Inversion}), which is typically accomplished
by a ridge extraction method that determines the maxima in the time-frequency
plot $|S_{\epsilon}^{\delta,M}f(t,\eta)|$. A discretized formulation
of the steps (\ref{CWT})-(\ref{Inversion}) and related computational
details can be found in \cite{TBFW2013}.\\

The main concepts behind Synchrosqueezing can also be applied to other
underlying time-frequency representations. The paper \cite{TW11}
develops a parallel approach based on the short-time Fourier transform
(STFT), which is shown to have some advantages.%
\footnote{We present a slightly different formulation of the transform than
\cite{TW11} that is more comparable with the approach in \cite{DLW11}.%
} The \textit{STFT Synchrosqueezing} process is similar to the above
development, but instead of (\ref{CWT}) is based on the \textit{modified
STFT} for an appropriate window function $G$,
\begin{equation}
V_{G}f(t,z)=\int_{-\infty}^{\infty}f(u)G(u-t)e^{-2\pi iz(u-t)}du.\label{mSTFT}
\end{equation}

This is simply the standard STFT with an additional modulation factor
$e^{2\pi izt}$, and can be thought of as a filter bank taken by sliding
the window $G$ over different frequency bands. The phase transform
(\ref{PhaseTransform}) and Synchrosqueezing transform (\ref{SST})
respectively become
\[
\tilde{\omega}f(z,t)=\frac{\frac{\partial}{\partial t}V_{G}f(t,z)}{2\pi iV_{G}f(t,z)},
\]
\begin{equation}
\tilde{S}_{\epsilon}^{\delta,M}f(t,\eta)=\int_{\{(t,z):z\in[M^{-1},M],|V_{G}f(t,\eta)|>\epsilon\}}V_{G}f(t,z)\frac{1}{\delta}h\left(\frac{\eta-\tilde{\omega}f(t,z)}{\delta}\right)dz,\label{STFT-SST}
\end{equation}
and the components can be recovered by fully inverting (\ref{STFT-SST})
as before by taking
\begin{equation}
\tilde{R}_{k,\epsilon}^{\delta,M}f(t)=\frac{1}{\int_{-\infty}^{\infty}|G(z)|^{2}dz}\int_{|\eta-\phi_{k}'|<\epsilon}\tilde{S}_{\epsilon}^{\delta,M}f(t,\eta)d\eta.\label{Inversion-STFT}
\end{equation}

\begin{figure}[h]
\begin{centering}
\includegraphics[scale=0.5]{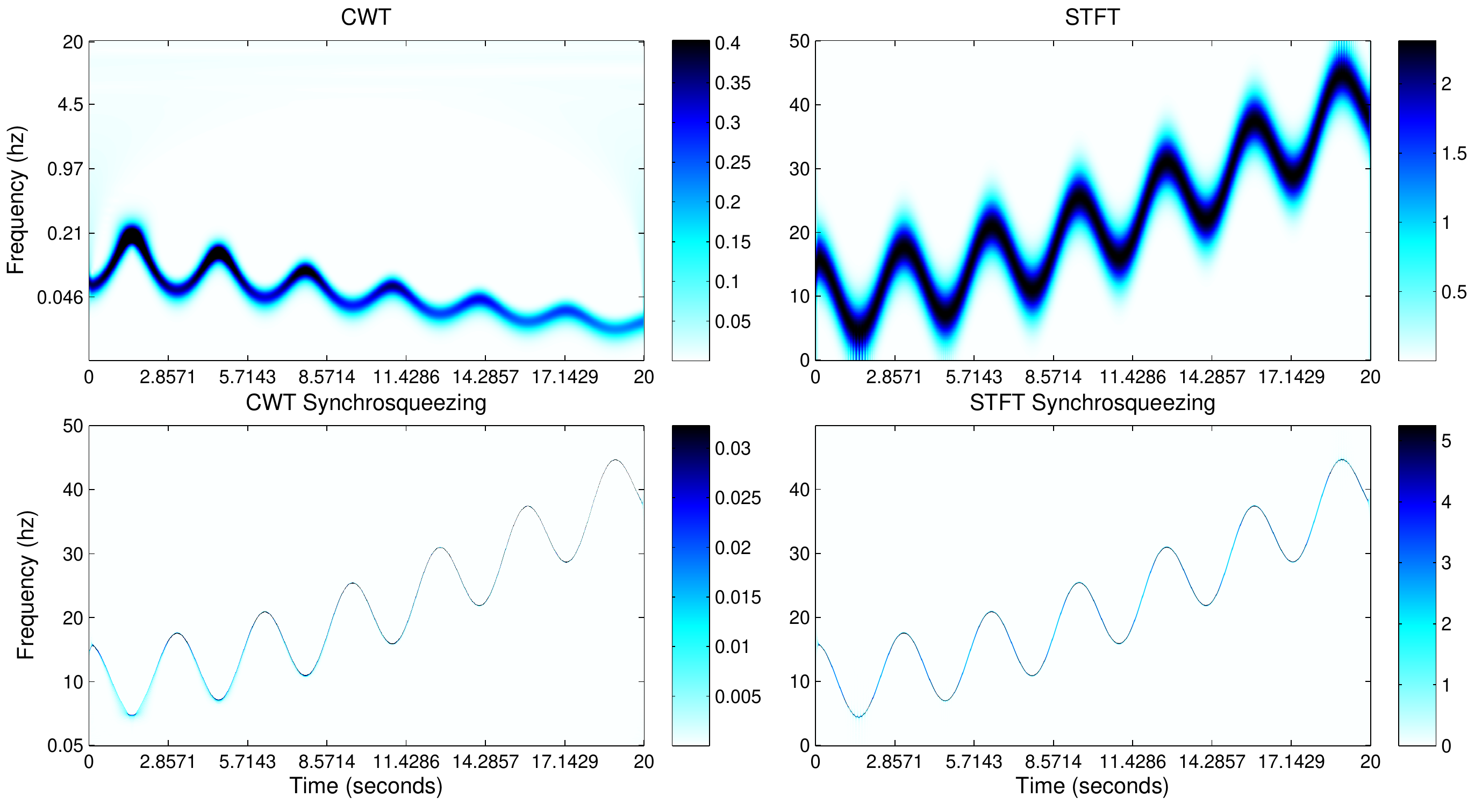}
\par\end{centering}

\centering{}\caption{\label{FigEx} Time-frequency plots of the signal $f(t)=\cos(2\pi(0.1t^{2.6}+3\sin(2t)+10t))$
under different transforms.}
\end{figure}

A simple example of the time-frequency plots $|S_{\epsilon}^{\delta,M}(t,\eta)|$
and $|\tilde{S}_{\epsilon}^{\delta,M}f(t,\eta)|$ is shown in Figure
\ref{FigEx}. While the traditional STFT and CWT plots are blurry,
reflecting the fact that they are not sparse representations of the
signal, the Synchrosqueezing transforms have a much more concentrated
profile and distinct IF curves in the time-frequency plane. Several
additional examples can be found in \cite{TBFW2013}, comparing CWT
Synchrosqueezing with TFR methods and other techniques. An open source
MATLAB toolbox implementing both forms of Synchrosqueezing is available
\cite{SSToolbox} and has facilitated the use of the technique across
different disciplines.\\

We briefly describe several extensions of these concepts that have
been developed. The paper \cite{Wu12} considers a variant of the
signal model in (\ref{AMFM}), where the mode $A_{k}(t)e^{2\pi i\phi_{k}(t)}$
is replaced by a more general form $A_{k}(t)s(\phi_{k}(t))$ for a
given ``shape function'' $s$ chosen to fit a particular application
at hand. This turns out to be a natural model for the analysis of
electrocardiogram signals, in which the sharp spikes (see Figure \ref{FigECG})
are not well represented by standard Fourier harmonics. In \cite{LL12b},
another generalization is presented based on replacing (\ref{mSTFT})
with a ``generalized Fourier transform,'' i.e an oscillatory integral
of the form $\int_{-\infty}^{\infty}f(u)g(u-t)e^{-2\pi iz\theta(u)}du$
where $\theta$ is a nonlinear phase function incorporating some prior
knowledge of the signal's structure. The paper \cite{Ya13} develops
another approach based on wave packet transforms, which encompasses
some aspects of both the CWT and STFT formulations.

\section{Theory}

Synchrosqueezing has a fairly comprehensive mathematical theory developed
for it, providing performance guarantees on selected classes of signals.
As of 2014, most of the published theory in \cite{DLW11} and \cite{TBFW2013}
covers the CWT version (\ref{SST}), but analogous results can be
shown for the STFT formulation (\ref{STFT-SST}) from \cite{TW11}
using similar techniques. We review the results for the CWT case here,
which are based on a sparsity model for the signal (\ref{AMFM}) in
the frequency domain.
\begin{defn}
\label{AClass}For given parameters $\epsilon,d>0$, we define the
class $\mathcal{A}_{\epsilon,d}=\{f:f(t)=\sum_{k=1}^{K}A_{k}(t)e^{2\pi i\phi_{k}(t)}\}$,
where 
\begin{align*}
 & A_{k}\in L^{\infty}\cap C^{1},\quad\phi_{k}\in C^{2},\quad\phi'_{k},\phi''_{k}\in L^{\infty},\quad A_{k}(t)>0,\quad\phi_{k}'(t)>0\\
 & \forall t\quad\left|A_{k}'(t)\right|\leq\epsilon\left|\phi_{k}'(t)\right|,\quad\left|\phi_{k}''(t)\right|\leq\epsilon\left|\phi_{k}'(t)\right|,\quad\mathrm{and}
\end{align*}

\begin{align}
\frac{\phi'_{k}(t)-\phi'_{k-1}(t)}{\phi'_{k}(t)+\phi'_{k-1}(t)} & \geq d.\label{Separation}
\end{align}

\end{defn}
The key condition here is (\ref{Separation}), which says that higher
frequency IFs are spaced exponentially further apart than lower frequency
IFs. Under this signal model, the following result can be obtained
\cite{DLW11}.
\begin{thm}
\label{ThmSS}Let $f\in\mathcal{A}_{\epsilon,d}$ for some $\epsilon,d>0$,
$h\in C_{0}^{\infty}$ with $\left\Vert h\right\Vert _{L^{1}}=1$,
and $\psi\in C^{1}$ with $\hat{\psi}$ supported in $[1-\Delta,1+\Delta]$
for some $\Delta<\frac{d}{1+d}$. Let $M$ be sufficiently large and
define $\tilde{\epsilon}=\epsilon^{1/3}$ and the ``scale band''
$Z_{k}=\{(a,b):|a\phi_{k}'(t)-1|<\Delta\}$. If $(a,t)\in Z_{k}$
and $|W_{\psi}f(a,t)|>\tilde{\epsilon}$, then $|\omega f(a,t)-\phi_{k}'(t)|\leq\tilde{\epsilon}$.
Conversely, if $(a,t)\not\in Z_{k}$ for any $k$, then $|W_{\psi}f(a,t)|\leq\tilde{\epsilon}$.
Futhermore, for some constant $C_{1}$,
\[
\left|\lim_{\delta\rightarrow0}R_{k,\tilde{\epsilon}}^{\delta,M}f(t)-A_{k}(t)e^{2\pi i\phi_{k}(t)}\right|\leq C_{1}\tilde{\epsilon}.
\]

\end{thm}
This result says that the energy in the Synchrosqueezing time-frequency
plane is concentrated around the IF curves $\{\phi_{k}'(t)\}$, and
the inverted components $f_{k}$ approximate the actual oscillatory
components $\{A_{k}e^{2\pi i\phi_{k}}\}$. Additional results of this
type were proved in \cite{TBFW2013}, describing the robustness of
the Synchrosqueezing transform under unstructured perturbations (e.g.
quantization error) as well as white noise. We slightly paraphrase
these theorems for clarity.
\begin{thm}
\label{SSStable}Let $f$, $\epsilon$, $d$, $h$, $\psi$ and $\Delta$
be given as in Theorem \ref{ThmSS}. Let $g=f+E$ for some error $E$
with $\left\Vert E\right\Vert _{L^{\infty}}$ sufficiently small.
There are positive constants $M$, $C_{2}$, $C_{3}$ and $C_{4}$
such that the following holds. Let $a\in[\frac{1}{M},M]$. If $(a,t)\in Z_{k}$
and $|W_{\psi}g(a,t)|>C_{2}\tilde{\epsilon}$, then $|\omega g(a,t)-\phi_{k}'(t)|\leq C_{3}\tilde{\epsilon}$.
Conversely, if $(a,t)\not\in Z_{k}$ for any $k$, then $|W_{\psi}g(a,t)|\leq C_{2}\tilde{\epsilon}$.
Futhermore,
\[
\left|\lim_{\delta\rightarrow0}R_{k,C_{2}\tilde{\epsilon}}^{\delta,M}g(t)-A_{k}(t)e^{2\pi i\phi_{k}(t)}\right|\leq C_{4}\tilde{\epsilon}.
\]

\end{thm}
$\quad$
\begin{thm}
Let $f$, $\epsilon$, $d$, $h$, $\psi$ and $\Delta$ be given
as in Theorem \ref{ThmSS}, with $\psi$ also satisfying $\left|\left\langle \psi,\psi'\right\rangle \right|<\left\Vert \psi\right\Vert _{L^{2}}\left\Vert \psi'\right\Vert _{L^{2}}$.
Let $g=f+N$, where $N$ is Gaussian white noise with power $\epsilon^{2+p}$
for some $p>0$. There are positive constants $M$, $E_{1}$, $E_{2}$,
$C_{2}'$, $C_{3}'$ and $C_{4}'$ such that the following holds.
Let $a\in[\frac{1}{M},M]$. If $(a,t)\in Z_{k}$ and $|W_{\psi}g(a,t)|>C_{2}'\tilde{\epsilon}$,
then with probability $1-e^{-E_{1}\epsilon^{-p}}$, $|\omega g(a,t)-\phi_{k}'(t)|\leq C_{3}'\tilde{\epsilon}$.
Conversely, if $(a,t)\not\in Z_{k}$ for any $k$, then with probability
$1-e^{-E_{2}\epsilon^{-p}}$, $|W_{\psi}g(a,t)|\leq C_{2}'\tilde{\epsilon}$.
Futhermore, with probability $1-e^{-E_{1}\epsilon^{-p}}$, 
\[
\left|\lim_{\delta\rightarrow0}R_{k,C_{2}\tilde{\epsilon}}^{\delta,M}g(t)-A_{k}(t)e^{2\pi i\phi_{k}(t)}\right|\leq C_{4}'\tilde{\epsilon}.
\]

\end{thm}
For STFT Synchrosqueezing, a result similar to Theorem \ref{ThmSS}
was proved in \cite{TW11}, although presented in slightly different
terms there. The main distinction with the STFT approach is that the
theory is developed for a different function class $\mathcal{B}_{\epsilon,d}$,
defined in the same way as $\mathcal{A}_{\epsilon,d}$ in Definition
\ref{AClass} but with (\ref{Separation}) replaced by the weaker
separation requirement that $\inf_{t}\phi_{k}'(t)-\sup_{t}\phi_{k-1}'(t)>d$.
The linear frequency scale of the modified STFT effectively allows
the IF curves $\{\phi_{k}'\}$ to be spaced much closer to each other
than the logarithmic scale of the CWT. In practical terms, STFT Synchrosqueezing
is well suited for decomposing signals with multiple components that
have closely packed IFs, especially at higher frequencies, while CWT
Synchrosqueezing is more appropriate for studying low frequency, trend-like
components in a signal.\\

We finally mention that the above results have mostly been formulated
in a deterministic setting, where the signal of interest $f$ is assumed
to lie in the class $\mathcal{A}_{\epsilon,d}$ but without any particular
mechanism that generated it. The paper \cite{CCW13} develops extensions
of these ideas to a stochastic model of the form $Y(t)=f(t)+T(t)+X(t)$,
where $f$ is essentially of the type $\mathcal{A}_{\epsilon,d}$,
$T$ is a slowly varying trend and $X$ is an autoregressive moving
average (ARMA) process with a time-dependent variance. The authors
use CWT Synchrosqueezing to extract the components $f$, $T$ and
$X$ from an observed signal $Y$, and prove several results on confidence
bounds and other aspects of the decomposition.

\section{Applications}

\begin{figure}[H]
\begin{centering}
\includegraphics[width=0.78\textwidth]{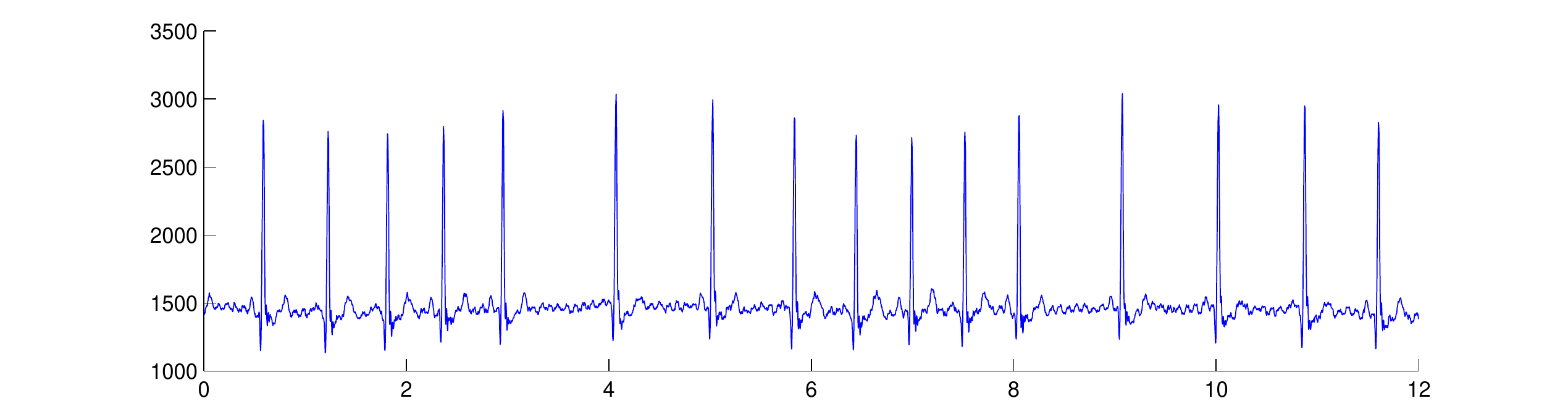}
\par\end{centering}

\centering{}\includegraphics[width=0.78\textwidth]{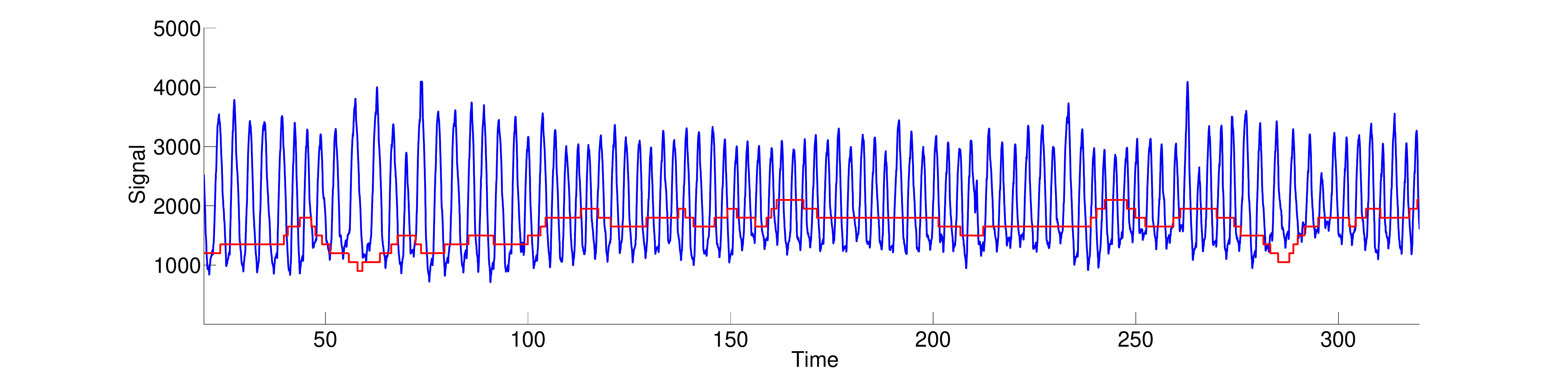}
\caption{\label{FigECG}Top: 10 second portion of ECG signal. Bottom: True
respiration signal (blue) and the IF computed from the ECG signal's
R peaks (red) using STFT Synchrosqueezing.}
\end{figure}

Due to its wide applicability, the Synchrosqueezing transform has
been used to address problems in many diverse disciplines. The technique
was first applied to topics in cardiology, specifically the analysis
of electrocardiogram (ECG) signals \cite{TW11,WCL14,WHB14}. The sharp
spikes in an ECG signal are called the \textit{R peaks} (see Figure
\ref{FigECG}) and encode important information about a patient's
heart rate, respiration and many other physiological properties. The
analysis of respiration, or breathing characteristics, is important
in many clinical applications such as testing for sleep apnea. However,
recording the respiration directly requires hooking up a breathing
apparatus (ventilator) to the patient and is often impractical to
perform over a long period of time. A patient's respiration influences
the ECG measurement and can be modeled as a low frequency envelope
fitting over the R peaks, with the ECG signal's IF closely following
the unobserved respiration signal's IF. The R peaks are not spaced
uniformly but can be used to form an impulse train $\sum_{k}f(t_{k})\delta(\cdot-t_{k})$,
where $\{t_{k}\}$ are the locations of the R peaks. Applying the
STFT Synchrosqueezing transform to this impulse train provides an
IF that accurately reflects short-range frequency variations in the
respiration signal (Figure \ref{FigECG}), and can be used for diagnosing
irregularities in the patient's breathing.\\

\begin{figure}[H]
\centering{}\includegraphics[width=0.52\textwidth]{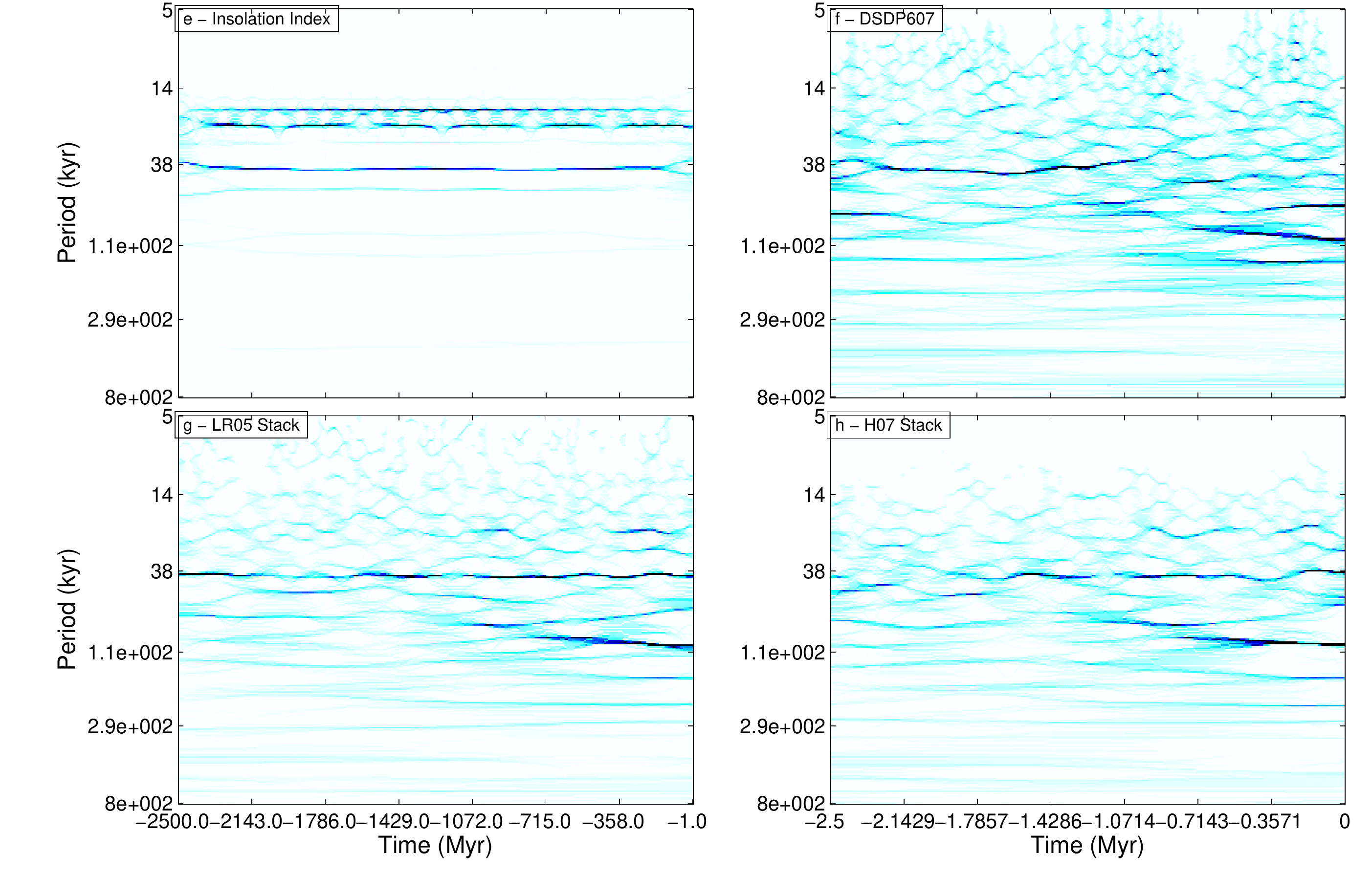} \includegraphics[width=0.45\textwidth]{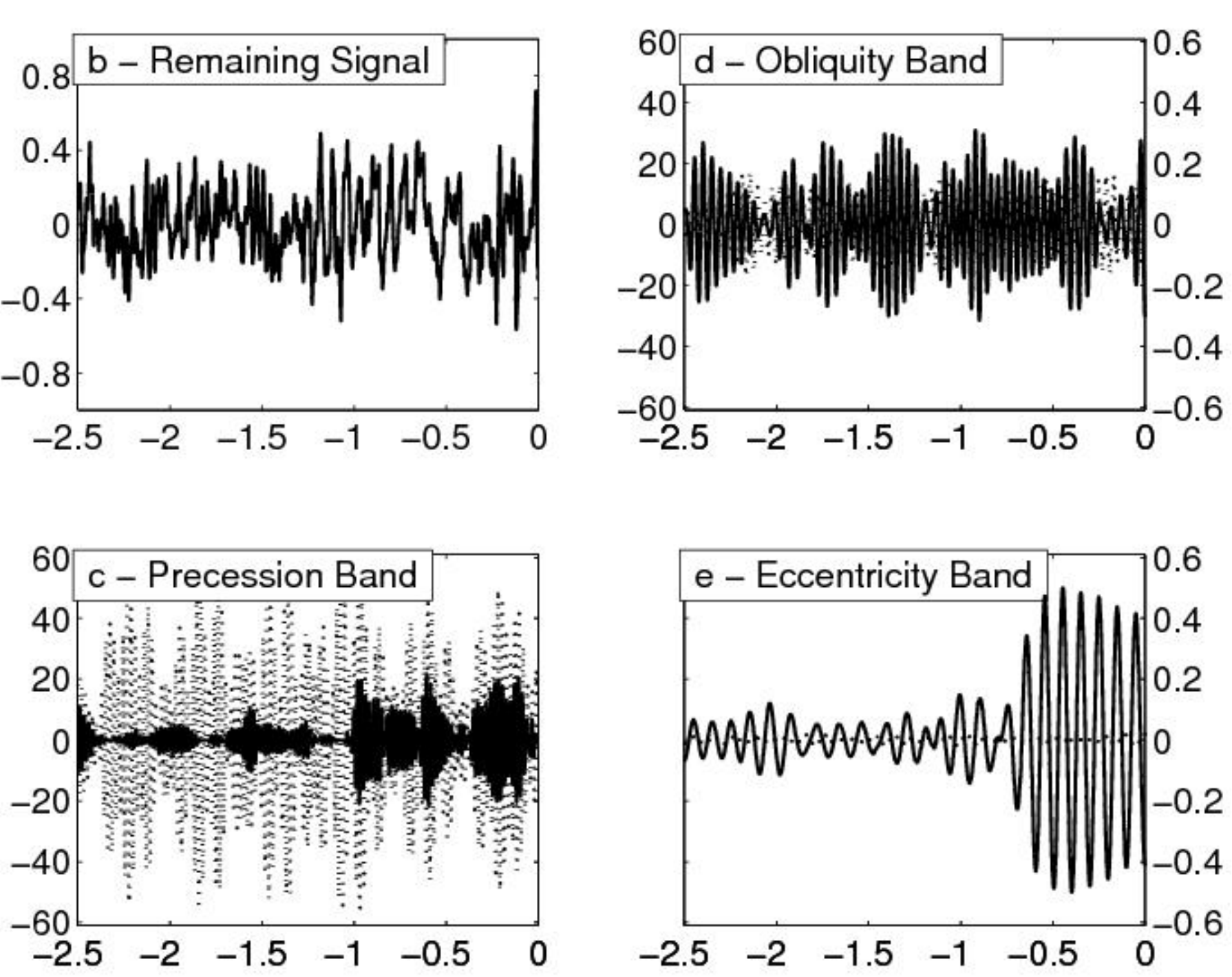}\caption{\label{FigClimate}Left: CWT Synchrosqueezing plots of the insolation
index, a single core (DSPD07) and stacks of such cores (LR05 and H07).
Right: Reconstructed oscillatory components, corresponding to the
obliquity, precession and eccentricity cycles.}
\end{figure}

Synchrosqueezing has also been used for the analysis of long term
trends in the global climate. The paper \cite{TBFW2013} studies sediment
cores extracted from the ocean floor, in which the relative concentrations
of the oxygen isotopes $\delta^{18}O$ and $\delta^{16}O$ indicate
changes in the sea level, ice volume and deep ocean temperature. These
are caused by long term fluctuations in the Earth's eccentricity and
other rotational properties over time, known as \textit{Milankovitch
cycles}, which influence the amount of solar radiation received at
the top of the atmosphere. The CWT Synchrosqueezing transform is used
to analyze the $\delta^{18}O$ levels in several composite stacks
of cores over the last 2.5 million years (Figure \ref{FigClimate}).
It is able to distinguish the different Milankovitch cycles more accurately
than the regular CWT, commonly used in this field, and identify when
certain components faded away or became more prominent. The invertibility
of the transform also allows one to extract the oscillatory components
corresponding to each of the Milankovitch cycles, and better characterize
some sudden changes in the climate between 0.5 and 1 million years
ago.\\

\begin{figure}[H]
\centering{}\includegraphics[clip,scale=0.4]{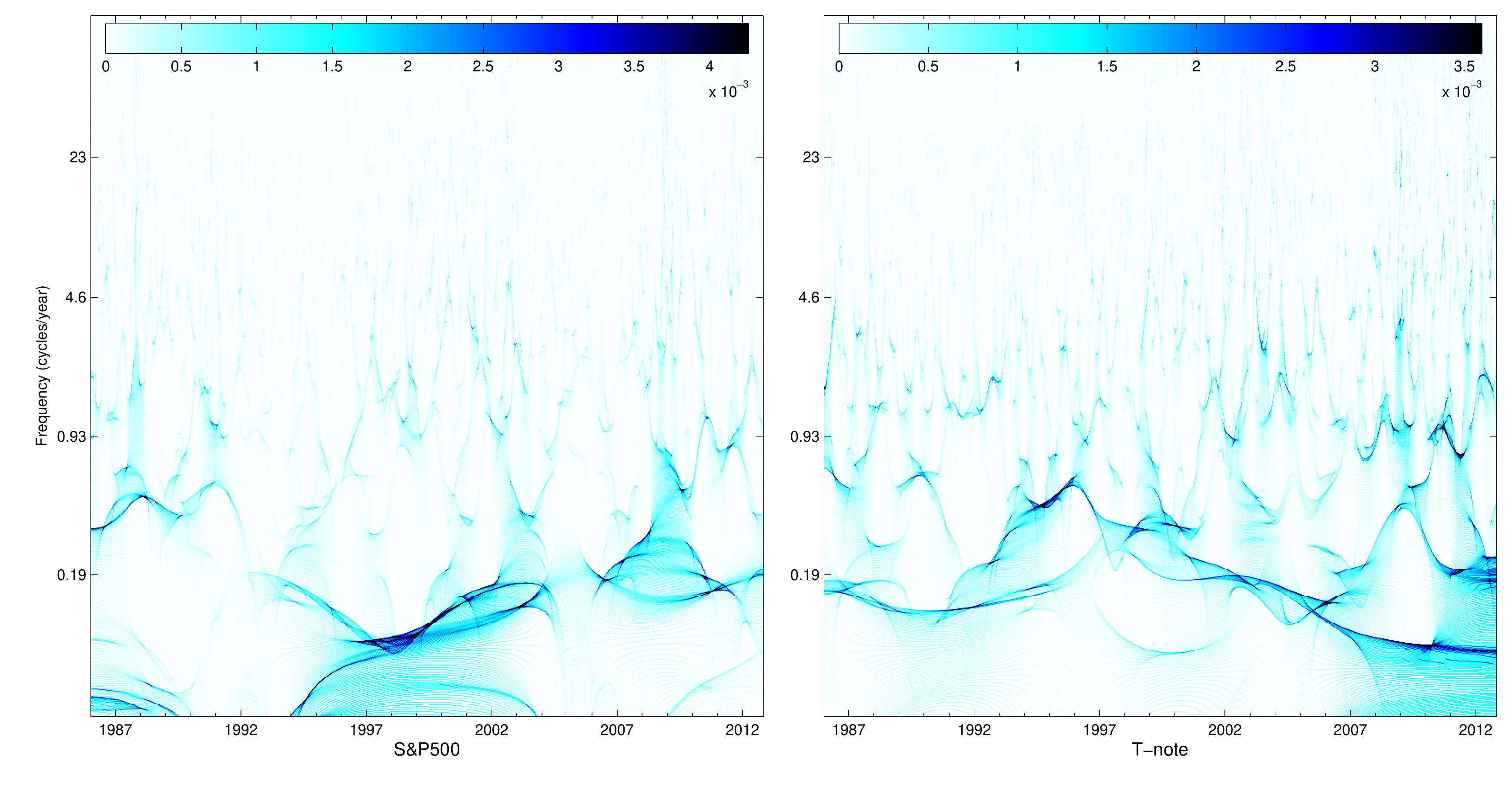}\caption{\label{FigEcon}CWT Synchrosqueezing plots of the S\&P 500 price and
the 10-year US Treasury yield.}
\end{figure}

Another application of Synchrosqueezing can be found in economics.
The paper \cite{GTG13} studies the stability of the US financial
system by considering time-frequency decompositions of equity indices,
Treasury yields, foreign exchange rates and several other macroecomonic
time series. Each time series is thought of as the output of a dynamical
system that produces slowly time-varying frequencies of the form (\ref{AMFM}),
but which are interspersed by abrupt frequency transitions (\textit{structural
breaks}) that indicate the starting or stopping of new underlying
dynamics. Among other events, the stock market crash in 1987 is contrasted
with the global recession in 2008. It is shown that the former had
a minimal impact on the dominant, low frequency components despite
being prominent in the original data, while the latter was both preceded
and followed by a variety of new dynamics, which left the economy
in a permanently altered state (Figure \ref{FigEcon}). The authors
also discuss a measure of instability in a time series called the
``density index,'' taking the $L^{1}$ norm of the IFs at each point
in time as a measure of how spread out or concentrated the frequencies
are. A sharp jump in the density index corresponds to a structural
break, which is shown to coincide with some of the major financial
stress events over the last 25 years and which may provide ``early
warning'' signs of future economic crises.\\

We briefly mention several other applications of Synchrosqueezing
that have appeared in the literature. In \cite{LL12}, it is used
to detect and analyze faults in a mechanical gearbox. The Synchrosqueezing
plot of the gearbox's vibration signal reveals extra sideband components
surrounding a central IF curve, which indicate the presence of a chipped
gear in the transmission. In geophysics, \cite{HTB13} discusses the
use of Synchrosqueezing to separate out resonant frequencies in data
from micro-seismic experiments, which are used to study deformations
in injection wells for oil extraction. Finally, \cite{ATM12} develops
an automated trading strategy based on Synchrosqueezing, using the
technique to model the relationship between correlated asset pairs
such as the stocks of competing firms. The rise in one asset's price
often precedes a fall in the other one, and a strategy based on identifying
the prices' IFs is shown to describe short-range oscillations and
outperform some standard approaches used in the industry.


\bibliographystyle{plain}
\bibliography{Fullbib}

\end{document}